%% file: draft-IEEE.tex
\documentclass[conference]{IEEEtran}
\IEEEoverridecommandlockouts
\input{macros}
\usepackage{flushend}

\begin{document}
\title{A Privacy-Preserving Longevity Study of Tor's Hidden Services}
% \titlenote{Produces the permission block, and
% \subtitle{Extended Abstract}
% \subtitlenote{The full version of the author's guide is available as
%   \texttt{acmart.pdf} document}

\author{
    \IEEEauthorblockN{Amirali Sanatinia\IEEEauthorrefmark{1}, Jeman Park\IEEEauthorrefmark{2}, Erik-Oliver Blass\IEEEauthorrefmark{3}, Aziz Mohaisen\IEEEauthorrefmark{2}, Guevara Noubir\IEEEauthorrefmark{1}}
    \IEEEauthorblockA{\IEEEauthorrefmark{1}Northeastern University
    \\\{amirali, noubir\}@ccs.neu.edu}
    \IEEEauthorblockA{\IEEEauthorrefmark{2}University of Central Florida
    \\\{parkjeman, mohaisen\}@cs.ucf.edu}
    \IEEEauthorblockA{\IEEEauthorrefmark{3}Airbus
    \\\{erik-oliver.blass@\}@airbus.com}
}

% \author{\IEEEauthorblockN{Amirali Sanatinia}
% \IEEEauthorblockA{Northeastern University\\
% amirali@ccs.neu.edu}

% \and
% \IEEEauthorblockN{Jeman Park}
% \IEEEauthorblockA{University of Central Florida\\
% parkjeman@knights.ucf.edu}

% \and
% \IEEEauthorblockN{Erik-Oliver Blass}
% \IEEEauthorblockA{Airbus\\
% erik-oliver.blass@airbus.com}

% \and
% \IEEEauthorblockN{Aziz Mohaisen}
% \IEEEauthorblockA{University of Central Florida\\
% mohaisen@cs.ucf.edu}

% \and
% \IEEEauthorblockN{Guevara Noubir}
% \IEEEauthorblockA{Northeastern University\\
% noubir@ccs.neu.edu}
% }

% \renewcommand{\shortauthors}{B. Trovato et al.}

\maketitle

\input{abstract}
\input{intro}

\input{tor}
\input{simulation}
\input{threatmodel}
\input{privcount}
\input{protection}

\input{implementation}
\input{results}

\input{related}
\input{conclusion}

\bibliographystyle{abbrv}
\bibliography{bibliography}

\end{document}

%% file: macros.tex
\usepackage{amssymb,amsmath}
\usepackage[linesnumbered,boxed]{algorithm2e}
\usepackage{epsfig}
\usepackage{fixltx2e}
\usepackage{url}
\usepackage{graphicx}
\usepackage{subfig}
\usepackage{natbib}
\setcitestyle{numbers,sort&compress}

\newcommand{\BfPara}[1]{{\vskip 1eX\noindent {\bf #1}}}
\newcommand{\ignore}[1]{}

\newcommand{\Dis}[0]{\mathsf{D}}
\newcommand{\histo}[0]{\mathsf{histogram}}

\newcommand{\getr}[0]{\stackrel{\$}{\leftarrow}}

\newcommand{\Z}[0]{\mathbb{Z}}
\newcommand{\myS}[0]{\mathcal{S}}

\makeatletter
\def\old@comma{,}
\catcode`\,=13
\def,{%
  \ifmmode%
    \old@comma\discretionary{}{}{}%
  \else%
    \old@comma%
  \fi%
}
\makeatother

%% file: abstract.tex
\begin{abstract}

  Tor and hidden services have emerged as a practical solution to
  protect user privacy against tracking and censorship. At the same
  time, little is known about the lifetime and nature of hidden
  services. Data collection and study of Tor hidden services is
  challenging due to its nature of providing privacy.  Studying the
  lifetime of hidden services provides several benefits. For example,
  it allows investigation of the maliciousness of domains based on
  their lifetime. Short-lived hidden services are more likely not to
  be legitimate domains, e.g., used by ransomware, as compared to
  long-lived domains.

  In this work, we investigate the lifetime of hidden services by
  collecting data from a small (2\%) subset of all Tor HSDir relays in
  a privacy-preserving manner. Based on the data collected, we devise
  protocols and extrapolation techniques to infer the lifetime of
  hidden services. Moreover we show that, due to Tor's specifics, our
  small subset of HSDir relays is sufficient to extrapolate lifetime
  with high accuracy, while respecting Tor user and service privacy
  and following Tor's research safety guidelines. Our results indicate
  that a large majority of the hidden services have a very short
  lifetime. In particular, 50\% of all current Tor hidden services
  have an estimate lifetime of only 10 days or less, and 80\% have a
  lifetime of less than a month.
\end{abstract}

%% file: intro.tex
% \fixme{Stress efficiency improvement over paper by Sherr.}
\section{Introduction}\label{s:intro}
Only little is known about Tor's hidden services, their nature, and
their lifetime. Yet, studying the lifetime (``longevity'') of hidden
services can provide several benefits: for example knowledge about
service lifetime provides insights into the performance and resource
allocation requirements of Tor's infrastructures, knowledge about the
typical lifetime might be an indicator of maliciousness of individual
onion domains. Short-lived hidden services are considered more likely
not to be legitimate domains, as compared to long-lived domains.

The distributed, anonymity-providing nature of Tor renders any study
about hidden services non-trivial and technically challenging. To
enable users to connect to hidden services with anonymity, Tor employs
a complex rendezvous protocol. Hidden services announce themselves on
a special subset of all Tor relays, the so called Hidden Service
Directory (HSDir) relays. These relays, currently around $3,000$,
store service descriptors uploaded by each hidden service and reply to
user connection requests in a distributed fashion.  Specifically, each
hidden service registers itself every day to a (pseudo-)randomly
chosen set of 6 HSDir relays. Only these 6 relays know about the
hidden service and user connection requests.  If the party conducting
the study would have access to all HSDir relays, a lifetime study of
hidden services would become trivial.  The party would simply count
the number of days a hidden service descriptor is uploaded to any
HSDir relay. However, given the decentralized, volunteer-run nature of
Tor relays, such a straightforward approach is obviously
impossible. There will always be HSDir relays outside the measuring
party's access, resulting in an incomplete view of hidden services.

Besides such a purely technical challenge, there exist also ethical
challenges when conducting a study on Tor.  Any study collecting
real-world data from Tor's relays must consider implications of data
collection, storage, and analysis.  Publishing (statistical)
information about real-world data and performing the study itself must
not de-anonymize individual parties.  Therefore, the recently
established Tor Research Safety Board reviews new research proposals
and approves only projects with minimal privacy risks and an expected
benefit for the Tor community.%~\cite{tor-research-safety-board}.

\vskip 1eX\noindent{\bf This paper:} We measure and analyze Tor's
hidden services lifetime  in a privacy-preserving
fashion. First, we investigate the effect of having access to only a
fraction of HSDir relays on the precision of hidden service lifetime
estimation. Surprisingly, we are able to show that access to even a
small fraction ($\approx{}2\%$) already allows lifetime prediction
with high precision, independent of the real distribution of
lifetimes.

We then institute several changes to the Tor relay software. First,
with our software, a relay counts the number of observations of each
individual hidden service descriptor and user connection requests to
this hidden service.  To protect anonymity of individuals and hidden
services, we also design and implement new techniques for multiple
different parties running multiple relays. At the end of a measuring
epoch, a lifetime  histogram is computed in a
privacy-preserving manner. Following the standard rationale of secure
multi-party computation, each (potentially malicious) party running a
set of relays only learns the information from their set of relays and
the histograms, but nothing else.  We store observation and request
counts with forward security. That is, if a relay is compromised by an
adversary at some point, the adversary will not learn count and
request values from before the time of compromise.

We have deployed a total of 80 modified relays on
cloud VMs, i.e., $2\%$ of all HSDir relays. Our set of relays has been
separated into different entities, based on the affiliation running
the relay, different cloud providers, and different geo-locations.
Our modified relays were monitoring hidden service and connection
requests over a period of 180 days.

This paper makes the following major {\bf contributions}:
\begin{itemize}

% \item \fixme{Formal relations between number of HSDir relays we run
%     compared to precision and actual distribution.}
  
\item We design and implement new analysis, simulation, and
  extrapolation techniques for Tor hidden service lifetime
  estimation. These techniques allows us to infer lifetimes with high
  accuracy using data collected from a small number of HSDir relays.

\item We design and implement a new protocol for multi-party,
  privacy-preserving computation of lifetime 
  histograms. Therewith, curious HSDir relays do not learn
  anything about hidden services besides what can be deduced from
  histograms. Our protocol is significantly more efficient than
  related approaches.

\item We deployed 80 relays over a period of 180 days. Our relays span
  over 3 different cloud platforms across USA and Europe (9 cities
  from 5 countries).
  %\fixme{Some %information about the deployment.}

\item As an outlook, we discuss several new techniques to cope with
  malicious adversaries which are dishonest regarding their input and
  thus violate correctness of a measurement. Our techniques are
  general and also apply to previous work on measuring Tor properties.
  
\end{itemize}

\vskip 1eX\noindent{\bf Remark:} We stress that technical details of
our approach were reviewed and approved by the Tor research safety
board before deployment.

%% file: tor.tex
\section{Background: Hidden Services \&\\ Hidden Service Directories}\label{s:tor}

\begin{figure*}[t!]
  \centering
  \begin{minipage}{0.49\textwidth}\centering
\includegraphics[height=3cm]{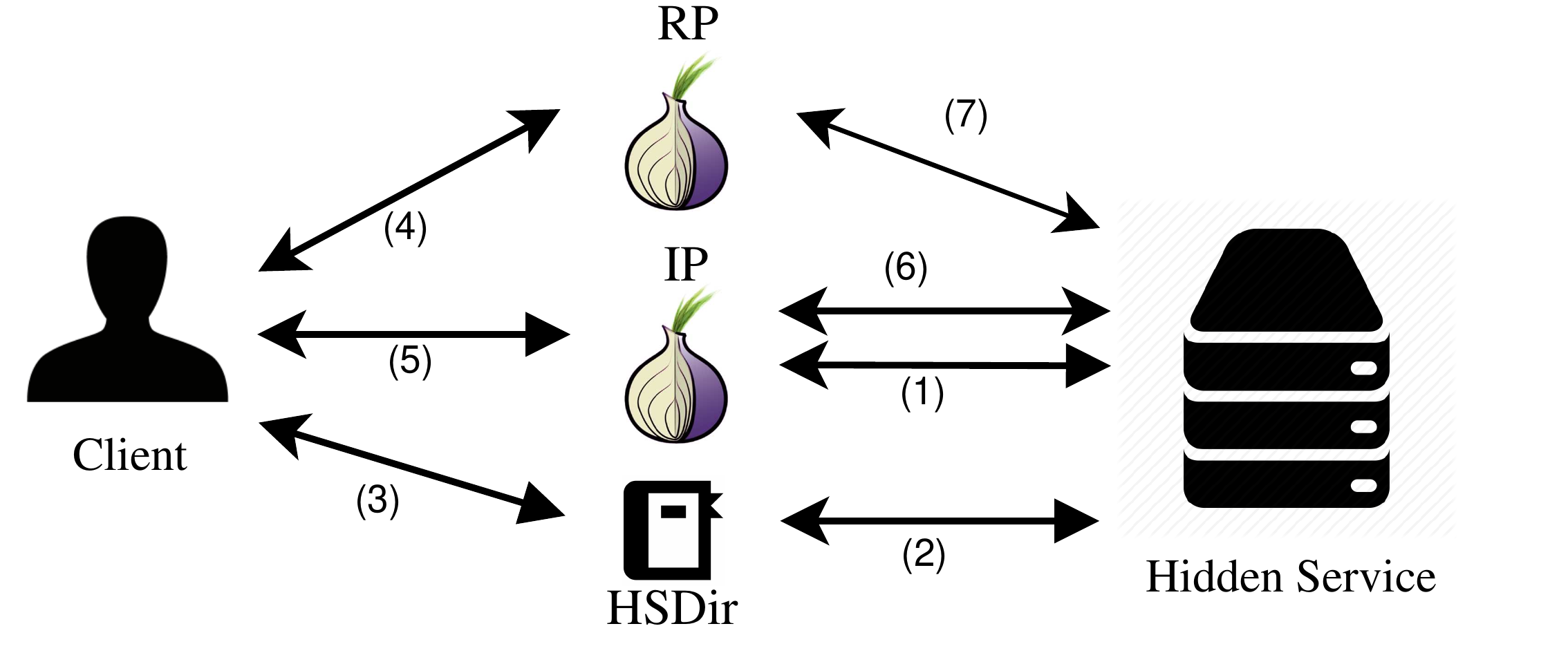}
\caption{\label{f:TorHS}Hidden service connection setup}
\end{minipage}
~
\begin{minipage}{0.49\textwidth}\centering
\includegraphics[height=3cm]{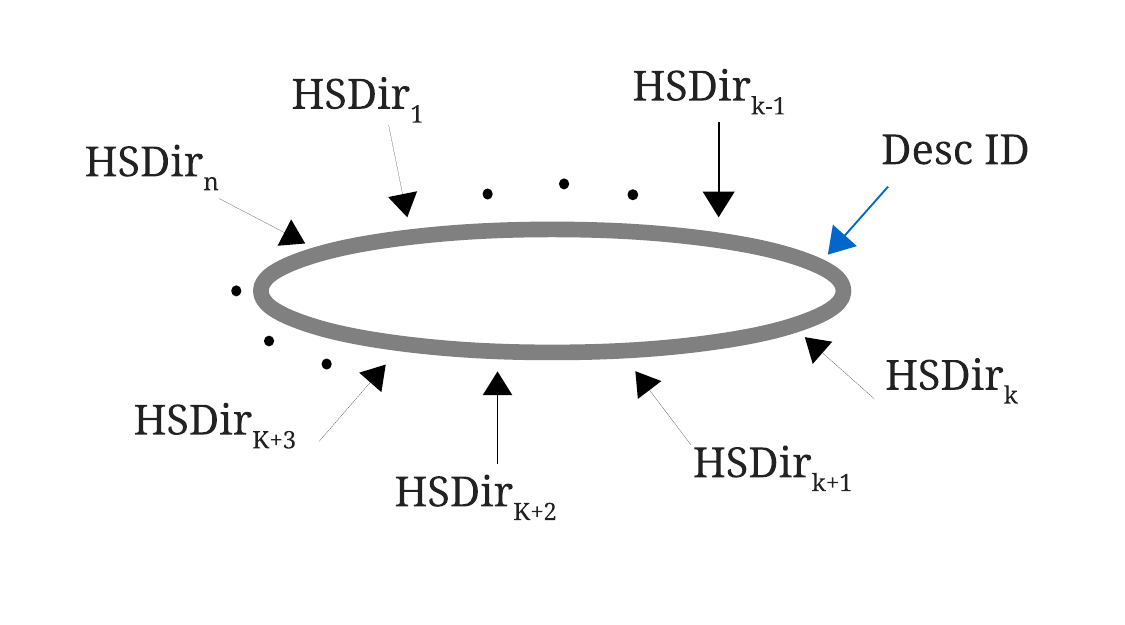}
\caption{\label{f:TorHSDir}
Ring of descriptor-ids and HSDir relay fingerprints}
\end{minipage}
\end{figure*}
Tor does not only anonymize client origin, but also network services
by introducing the concept of hidden services (also called onion
services). Roughly speaking, by connecting to a hidden service,
neither the client nor an eavesdropping adversary learn the service's
IP address and therewith location.  Hidden services have been used to
protect both legitimate, legal services for privacy conscious users
(e.g., Associated Press, CBC, or the Center for Public Integrity all
using SecureDrop through hidden services), but also for illicit
purposes such as drug and contraband markets and extortion.  We
briefly summarize key properties of Tor's hidden services.  In
particular, we focus on the architecture of hidden services, how
hidden services are offered such that clients and service providers
can communicate.  The Tor hidden services architecture is composed of
the following components:

\begin{itemize}
\item{\emph{Server}: computer running a service (e.g., a web server).}
\item{\emph{Client}: computer wishing to access the server.}
\item{\emph{Introduction Points (IP)}: a set of Tor relays, chosen by
  the hidden service server. They forward initial messages between the
  server and the client's Rendezvous Point.}
\item{\emph{Rendezvous Point (RP)}: a Tor relay randomly chosen by the
    client that will forward data between the client and the server.}
\item{\emph{Hidden Service Directories (HSDir)}: a set of Tor relays
  chosen pseudo-randomly to store certain service information.}
\end{itemize}

\BfPara{Server.}  The first step for the server is to make its service
accessible. Once for each service, the server generates an RSA key
pair. The (truncated) SHA-1 digest of the resulting public key is the
\texttt{Identifier} of the hidden service, and the \texttt{.onion}
hostname is the Base-32 encoding of this identifier. Typically, it is
sufficient for clients to know the \texttt{.onion} hostname to be able
to connect to the server over Tor. However before, the server
registers its hidden service inside Tor.  As shown in
Figure~\ref{f:TorHS}, the server randomly chooses a set of Introduction Points and establishes Tor circuits
with them (step 1). After setting up circuits, the hidden service then
calculates two {service \emph{descriptor-id}s and determines several
  HSDir relays, details below. The server uploads their list of
  Introduction Points to those HSDir relays (step 2).  The following
  simple relations hold for the descriptor-ids:
  \texttt{\small\hspace{-1mm}
\begin{eqnarray*}
  \mbox{descriptor-id} & = & \mbox{H(Identifier}||\mbox{secret-id-part)}\\
  \mbox{secret-id-part} & = &
 \mbox{H(time-period}||\mbox{descriptor-cookie} \\ && \mbox{||replica)}\\ 
 \mbox{time-period} & = & (\mbox{current-time} + \\
 && \mbox{permanent-id-byte}
 * 86400 / 256) \\ && / 86400
\end{eqnarray*}
}

Here, \texttt{H} denotes the SHA-1 hash function, \texttt{Identifier}
is the 80 bit truncated SHA-1 digest of the public key of the hidden
service. \texttt{Descriptor-cookie} is an optional 128 bit field which
could be used for additional authorization such that knowledge of the
\texttt{.onion} is not sufficient to connect to a hidden service.

Hidden services, i.e., their servers, change their HSDir relays
periodically.  The \texttt{time-period} determines when each
\texttt{descriptor-id} expires, and the hidden service needs to
calculate Introduction Points and uploads them to the new
corresponding HSDir relays. To prevent descriptor-ids from changing
all at the same time, a \texttt{permanent-id-byte} is also included in
the calculations. The \texttt{Replica} index takes values 0 or 1, and
thus there exist two descriptor-ids.

The range of all possible descriptor-ids forms a ring, see
Figure~\ref{f:TorHSDir}.  Additionally, each HSDir relay features a
fingerprint, the SHA-1 hash of their own public key.  These
fingerprints allowing placing HSDir relays on the ring of
descriptor-ids, essentially dividing the ring into segments.  Each
hidden service stores their Introduction Points with their
descriptor-id on three consecutive HSDir relays in the ring, so to a
total of six relays.  Specifically, if a descriptor-id of a hidden
service falls between fingerprints of HSDir relays HSDir$_{k-1}$ and
HSDir$_{k}$, then the Introduction Points will be stored on relays
HSDir$_{k}$, HSDir$_{k+1}$, and HSDir$_{k+2}$.

\BfPara{Client.} When a client wishes to connect to a hidden service,
they first need to compute the \texttt{descriptor-id} as above. The
client then contacts the corresponding HSDir relays (step
3) by establishing a circuit to receive the list of the hidden
service's Introduction Points. Then, the client first randomly selects
a Tor relay as its Rendezvous
Point and establishes a Tor circuit with it (step 4). The client
connects to one of the hidden service's Introduction Points and asks
to contact the hidden service, announcing their Rendezvous Point (step
5). Then, the Introduction Point forwards the Rendezvous Point to the
hidden services (step 6). Finally, the hidden service establishes a
circuit to the Rendezvous Point, and hidden service server and client
start communicating.

%% file: simulation.tex
\section{Measurement Configuration}\label{s:coverage}
%\noindent{\bf Rationale:}
Our key idea is to deploy several modified HSDir relays within the Tor
network to collect partial statistics. These HSDir relays follow the
standard Tor protocol specification with only one difference: the
relays maintain counters for all descriptor-ids observed. Whenever
\texttt{descriptor-id} is uploaded to one of our HSDir relays, this
relay will increase its local counter for \texttt{descriptor-id} by
one. After some time, which we divide into {\em epochs}, all of our
HSDir relays jointly produce a histogram of the hidden service
lifetime during that epoch. The histogram will capture how many hidden
services were observed on how many days. Computing the histograms
should be both precise and privacy-preserving. In this section, we
discuss our approach on how to measure with precision. We will present
the privacy-preserving computation aspects in the following section.

% Along the same lines, our HSDir relays will count client requests to
% access hidden service Introduction Points. Therewith, we will
% establish a simple hidden service popularity metric.\fixme{how was popularity by service computed in a privacy-preserving way?}

Deploying a large number of modified Tor relays such that nearly all
relays in the Tor network are our modified relays would obviously
result in high precision for the estimated lifetime and
popularity. However, deploying large numbers of relays clearly incurs
high (monetary) costs. As described later in
Section~\ref{s:longresults}, we use several different cloud providers
and deploy Tor relays on cloud VMs. The costs of running the
measurement is thus proportional to the number of relays we deploy.

Therefore, a first question is how many relays result in what level of
precision for our estimation.  In the following, we present our
approach to calculate the trade-off between the number of relays
(costs) on the one hand and the precision of lifetime and popularity
estimation on the other via simulation.

\subsection{Simulation Methodology}

We simulate the behavior of hidden services uploading their
descriptors to the HSDirs. As such, every hidden service chooses six
HSDir based on their fingerprint ring, following Tor's
specification. Given that we have the ground truth for the lifespan of
each hidden service, we are able to analyze the effect of the number
of controlled HSDirs on the statistical error, for each lifespan
$d$. Thus, the estimated lifespan $l_e(i)$ for $HS_i (i = 1, 2, \ldots
n)$ using the counts from all controlled HSDirs can be written as.
\[l_e(i) = \frac{\left(\sum_{k=1, 2 .. N_c}C_{k,i}\right) \times N_t}{N_c \times 6},\]
where $C_{k,i}$ is the counter value for the hidden service $HS_i$
(measured at $HSDir _k$), $N_c$ is the number of controlled HSDirs
(e.g., 80), and $N_t$ is the total number of HSDirs (e.g., 3,000) on
Tor. By dividing the sum of counters by $N_c$, we can estimate what
would have been obtained if we measured the results using all
HSDirs. During an epoch of 24 hours, six HSDirs store descriptors at
the same time, so we divide the results by six to estimate lifespan,
accounting for the over-count at those HSDirs.

\BfPara{Error Estimation.} The error for each hidden service, $E(i)$,
is calculated as the difference between the estimated and actual
lifespan. The total average error rate, $E_{avg}$ is calculated by
integrating these errors. Let $l_a(i)$ be the actual lifespan for
$HS_i$ from the initial phase. The error and total average error are
calculated as follows:
\begin{align}
    E(i) &= |l_e(i) - l_a(i)| / l_a(i),\\
    E_{avg}(N_c) &= [\sum_{i=1, 2..N_t}E(i)] / N_t.
\end{align}
Using a simulator we experiment with various variables to compute the above error rates and decide an acceptable measurement setting.

\BfPara{Simulation Setting.} During the simulation, we maintain the
total number of HSDirs in the network at 3,000 and the number of
hidden services at 60,000, which is the number of unique hidden
services per day~\cite{toruniquehs}. In addition, since the main
purpose of this simulation is to discover the correlation between the
number of controlled HSDirs and the error of the statistical results,
we do not consider users' behavior in the simulation. The number of
HSDirs controlled between 30 to 300, in steps of 30 (i.e., $N_c = 30,
60, 90, \dots, 300$).

\begin{figure}
     \centering
     \subfloat[][$N_C = 30$]{\includegraphics[width=0.48\textwidth]{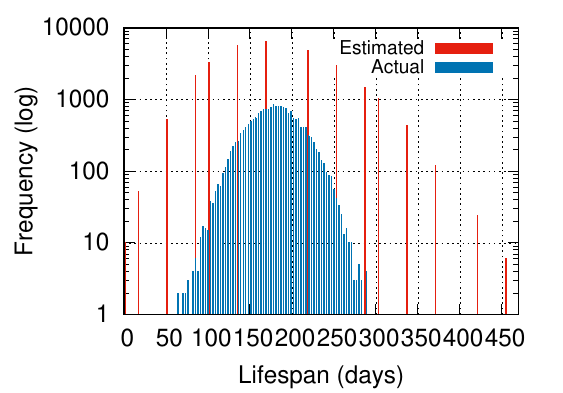}\label{fig:nc30}}\\
     \subfloat[][$N_C = 300$]{\includegraphics[width=0.48\textwidth]{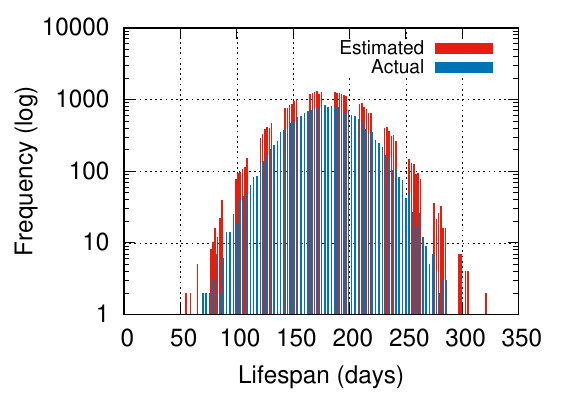}\label{fig:nc300}}
     \caption{The distributions of actual and estimated lifespan. Note that the distributions of actual lifespan in both cases are almost similar, but the distributions of estimated lifespan are quite different.}
     \label{fig:lifespanhistogram}
\end{figure}

\BfPara{Lifespan estimation using HSDirs.}
Figure~\ref{fig:lifespanhistogram} shows the distribution of the
actual and the estimated lifespan in our simulations with 30 and 300
controlled HSDirs. The actual lifespan (blue bars) is densely set
following a given normal random distribution, and the estimated
lifespan (red bars) looks rough as it is estimated from the counts
collected from the controlled HSDirs. Comparing the left and right
graphs, the actual lifespan of both graphs are observed to be similar,
but the estimated lifespan varies (per the variance or frequency of
the estimated values). This difference occurs because the range of
counter values to be collected can be widened when the number of the
controlled HSDirs is large and the range of the estimated lifespan
also widens together. In the right graph, the difference between the
estimated and actual lifespan is small, which means that the accuracy
of lifespan estimation is high.

\begin{figure}[htbp]
\centering
\includegraphics[width=0.45\textwidth]{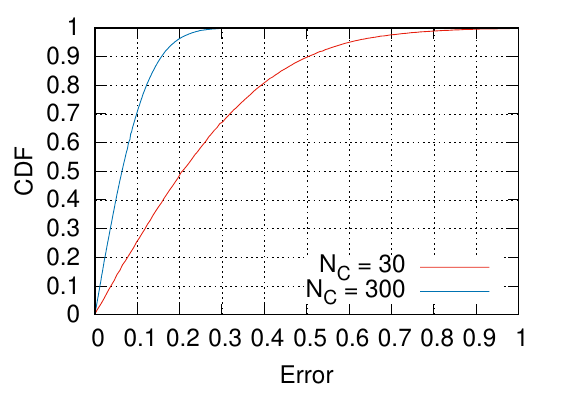}
\caption{CDF for hidden service errors. Note that when using 300
  HSDirs for the lifespan estimation only less than 10\% of HSs have
  an error over 0.2, while when using 30 HSDirs about half of HSs have
  error larger than 0.2}
\label{fig:CDF}
\end{figure}

Figure~\ref{fig:CDF} shows numerically a comparison between the
estimated and the actual lifespans. The X-axis represents the
calculated error of the hidden service using the above equation, while
the Y-axis is the number of hidden services that have the
corresponding error. In this graph, we can see that when $N_C$ is 300,
more than 90\% of HSs have an error of less than 0.2, while when $N_C$
is 30 only about 50\% have error of less than 0.2. This shows that the
accuracy of the accumulated results can vary depending on how many
HSDirs are used to collect information among all HSDirs.

\subsection{Determining the number of controlled HSDirs}

\begin{figure}[htbp]
\centering
\includegraphics[width=0.45\textwidth]{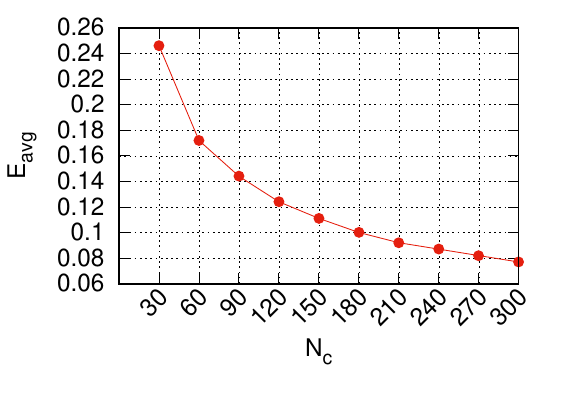}
\caption{The average expected error for various number of controlled HSDirs.}
\label{fig:error_comparison}
\end{figure}

The previous experiments have empirically demonstrated that the number
of controlled HSDirs can directly impact the measurement error. Based
on these observations, we measure how $ E_ {avg} $ varies according to
the number of HSDirs. It is clear that a large number of HSDirs
ensures high accuracy in the statistics, but it is necessary to find
an optimal value for realistic, safe, and practical measurements.

As shown in Figure~\ref{fig:error_comparison}, the average error rate
$E_{avg}$ changes as we we change the value of $N_c$. As the number of
the controlled HSDirs is increased, the estimated lifespan becomes
more accurate and the error rate decreases. Through this experiment,
we calculate the optimal number of controlled HSDirs to be between 60
and 90, as it is an appropriate compromise between cost and accuracy.

\subsection{Determining the number of contolled HSDirs for various distributions}

In the previous section, we investigated the optimal value of the
number of controlled HSDirs, assuming that the hidden services have a
lifespan following a normal distribution. However, assuming that it
follows a normal distribution in the absence of a baseline for the
true distribution of the lifespan of the hidden service introduces
additional errors. Therefore, we have also experimented with a case
where the lifespan of the hidden service is distributed along a
uniform and exponential distribution.

\begin{figure}[htbp]
\centering
\includegraphics[width=0.45\textwidth]{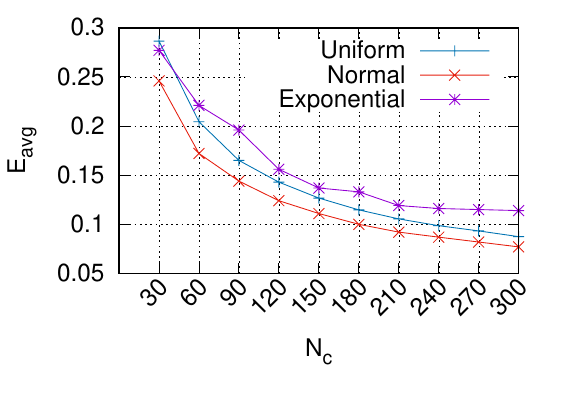}
\caption{The average of errors with the uniform, normal, and
  exponential distribution, respectively.}
\label{fig:error_distribution}
\end{figure}

Figure~\ref{fig:error_distribution} shows the average of errors of the
lifespan estimation with various random distributions. We note a
difference in the individual error rate, although the overall trend is
similar to our previous results. Through these extended experiments,
we also find that using the number of controlled HSDirs between 60 and
90 is an acceptable setting, regardless of the distribution of hidden
services' lifespan.

In the next section, we describe a privacy-preserving protocol that we
use to encrypt the lifespan counts for each hidden services, while
also hiding the raw hidden services addresses.

%% file: threatmodel.tex
\section{Privacy-Preserving Histograms}\label{s:threat}
We now turn to protecting privacy during our measurement study and
begin by discussing this paper's threat model. Essentially, our work
should not allow an adversary to learn more sensitive information than
if they would run several HSDir relays themselves.

\subsection{Threat Model}
As of today, a malicious party (an \emph{adversary}) running any HSDir
relay in Tor is able to get limited information about descriptor-ids
and client requests and therewith hidden services' lifetimes and
popularities.

Our experiment on the Tor network produces a histogram of hidden
service lifetimes. Informally, our privacy intuituion is that our
experiment must not enable the adversary to learn more than the
histograms and all data recorded by the adversary controlled HSDir
relays.

This should also hold for our own HSDir relays which we deploy inside
the Tor network using several cloud providers.  Some relays could be
compromised by an adversary. One could even argue that some members of
our team behaves adversarial (and thus we have separated control over
all relays to 3 different parties, see Section~\ref{s:impl}).

More formally, we assume that an adversary has compromised a subset
$\myS$ of our HSDir relays. Still, the adversary should not learn
anything besides
\begin{itemize}
\item what each compromised HSDir relay in $\myS$
  observes. Specifically, this comprises service descriptors and
  client requests.
  
  \item the histogram of lifetimes produced in the end.
  \end{itemize}

  An adversary either compromising some of our relays our
  simply running relays themselves can obviously combine the
  \emph{view} of their relays and possibly infer additional
  information. As the Tor network is based on volunteering, an
  adversary can participate at any time, and there is no way to
  prevent such kind of data leakage. The goal of our security
  mechanisms is to not introduce additional issues.

  As a side note we remark that an adversary compromising a relay
  might have full system access to the computer running the
  relay. This might include the power to dump the computer's memory,
  read swap data etc. Against such an adversary it is difficult to
  protect sensitive data, e.g., service descriptors, from before the
  time of compromise. To offer at least some security, we will use
  techniques to hide previously stored information.

%% file: privcount.tex
\subsection{A New Protocol for \\Privacy-Preserving Histograms}\label{s:privcount}
We now describe technical details of our protocol which computes
histograms in a privacy-preserving way.  The main setup of data
collection and aggregation is based on and extends the second variant
of the PrivEx~\cite{Elahi14} protocol.  PrivEx introduces intuitive
distributed encryption and multi-party computation for a single count
and has already served as the basis for other work,
too~\cite{Jansen:2016:SMT:2976749.2978310}.  In the following, we
present the main protocol steps. To avoid confusion, we use the same
notation and terminology (TKS and DC) as PrivEx.
% \fixme{highlight differences!}

\subsubsection{Overview}

We separate trust into multiple different, mutually distrusting
\emph{parties}.  Each party runs one Tally Key Server (TKS) and
several Data Collection (DC) node.  Each DC node in our setting is an
HSDir relay. During data collection, each DC node stores a count, the
number of times a hidden service has registered its service descriptor
at this DC node. To realize that, each DC manages a set of (key,
value) pairs, where the key is the hash of a .onion hostname, and the
value is the count for that .onion hostname. We store counts using
additively homomorphic encryption, see below.

After some time interval (an epoch), all DCs send their (key,value)
pairs to their corresponding TKS. A TKS homomorphically adds values
(counts) reported for the same key (hostname). This is done only for
safety reasons, such that not all measurements of a DC are in case
that DC crashes within our multiple-month study.

At the end of the whole data collection phase, i.e., after several
epochs, all TKS interactively compute a lifetime histogram with
privacy. In our setup with multiple distrustful parties, computation
 resembles standard cryptographic multi-party computation.

 \BfPara{Cryptography Overview.}  During initialization, our protocol
 uses a standard distributed key generation technique to generate a
 (private, public) key pair for an additively homomorphic encryption
 scheme. Here, the public key will be known to everybody, but not the
 private key.  The private key is split into shares, and each TKS
 knows one share of the private key.  To decrypt a ciphertext, all TKS
 need to cooperate and contribute to the decryption.

The DCs use this public key to encrypt the counts during the data
collection phase. We employ the additively homomorphic variant of
Elgamal encryption with plaintexts ``in the exponent''.  Thanks to the
additively homomorphic property, each TKS homomorphically adds counts
until the end of the data collection phase. During an aggregation
phase, all TKSs homomorphically add their individual sums to compute
encrypted aggregated sums. To separate sums for the corresponding
.onion service, each TKS shuffles the list of aggregated sums.

Finally, each individual sum is decrypted by each TKS contributing
with to each decryption with its share of the private key.

\ignore{ Specifically, for a safe prime

(private, public) key pair ($a$, $A = g^a$). Encryption of a message
  $m \in \mathbb{Z}_{q}$ and a random value $r \in \mathbb{Z}_{q}$ is
  $E_{A}(r; m) = (g^r, A^r \cdot h^m)$. In this scheme, $g$ and $h$
  are the generators of a group $\mathbb{G}$ of order $q$. Note that
  this scheme is an additively homomorphic encryption construct, where
  $D_a(E_{A}(r_1; m_1)) \cdot D_a(E_{A}(r_{2}; m_{2})) = D_a(E_{A}(r_1
  + r_2; m_1 + m_2))$.

$E_{A}(r_{1}; m_{1}) = (g^r_{1}, A^{r_1} \cdot h^{m_1})$

$E_{A}(r_{2}; m_{2}) = (g^{r_2}, A^{r_2} \cdot h^{m_2})$ 

$D_a(E_{A}(r_1; m_1)) \cdot D_a(E_{A}(r_{2}; m_{2})) = D_a(E_{A}(r_1 + r_2; m_1 + m_2))$ \\

 The decryption of cipher tuple ($C_1$, $C_2$) is, $D_a(C_1, C_2) =
 DL_h(C_2/{C_1}^a)$, which is based on the calculation of discrete
 logarithm. Given that the size of message space is small in our
 experiments, it can easily be calculated with brute-force method. A
 count value cannot be larger than the duration of the study,
 $d$. Even if we control all HSDirs in the Tor network, the total
 count cannot be larger than $6\cdot d$. For further optimization and
 more efficient calculation of discrete logarithm, other general
 approaches such as baby-step giant-step~\cite{Shanks}, Pollard's
 kangaroo algorithm~\cite{Pollard:2000:KMD:2815936.2815952}, or
 Pollard's rho algorithm~\cite{pollard:1978} can be used.  }
%ignore

We use a public bulletin board (PBB) for storage of all the
intermediary and final results of the computations that are accessible
to all DC and TKS nodes, to read from and write to. To mitigate
against misbehavior and data manipulation by the PBB, all data can be
either signed by the publishers, or the PBB can act as an append only
database or ledger. We use a private git data repository where each DC
and TKS have their own private key to publish their data.

%\noindent The privacy preserving data collection and aggregation, follows a set% of steps and phases that are outlined below. \\

%\BfPara{Security Rationale:} \fixme{multiple parties, each with one TKS and a set of DCs, we trust at least one party.}

After this overview, we now turn to technical details.

\subsubsection{Technical Details}
\hspace{1eX}
\BfPara{Initial Setup.}
Initially, all parties agree on an asymmetric key. We use Elgamal's
additively homomorphic variant, as distributed key generation for
Elgamal is straightforward. All operations below take place in a group
where the Decisional Diffie–Hellman assumption is believed to
hold, e.g., the group of quadratic residues $QR_p$ of $\Z_p$, where
$p=2\cdot{}q+1$ is a large safe prime. For any $a\getr{}QR_p$, $g=a^2$
is a generator of $QR_p$.

During an initial setup stage, parties jointly agree on a public key
using the following standard distributed key generation
technique. Each party $i$ picks a random value $a_i\getr\Z_q$ to be
used as its share of a joint private key. The share of the joint
public key is calculated, $A_{i} = g^{a_i}$. To mitigate against an
adversarial party, each party first publishes a hash-based commitment
to $A_i$. After all parties have published their commitments, they
open them to reveal $A_i$. An
alternative would be to commit to $A_i$ and publish the commitment together
with a zero-knowledge proof of knowledge of $a_i$. 

Next, the joint public key $A$ is calculated as $A = \prod_i{A_i}$. 

Encryption of a message $m \in \mathbb{Z}_{q}$ is $E_{A}(r,m) = (g^r, A^r \cdot h^m)$

for a random value $r \in \mathbb{Z}_{q}$ and two generators $g$ and
$h$. Note  the encryption's  additively homomorphism, as we have
%$$D_a(E_{A}(r_1, m_1)) \cdot D_a(E_{A}(r_{2}, m_{2})) = D_a(E_{A}(r_1 +
%r_2, m_1 + m_2)),$$ as we have
\begin{align*}
E_{A}(r_{1}, m_{1}) &= (g^{r_{1}}, A^{r_1} \cdot h^{m_1})
\\
E_{A}(r_{2}, m_{2}) &= (g^{r_2}, A^{r_2} \cdot h^{m_2}) 
\\
D_a(E_{A}(r_1, m_1)) \cdot D_a(E_{A}(r_{2}, m_{2})) &= D_a(E_{A}(r_1 + r_2, m_1 + m_2)) 
\end{align*}

Our additive variant of Elgamal encryption supports
re-encryption. That is, for a ciphertext
$C=(C_1,C_2)=(g^r,A^r\cdot{}h^m)$, we can re-encrypt to
$C'=(C_1^{r'},C_2\cdot{}A^{r'})$ for a randomly chosen
$r'\in \mathbb{Z}_{q}.$

Decryption of ciphertext tuple ($C_1$, $C_2$) is,
$$D_a(C_1, C_2) = DLog_h(C_2/{C_1}^a).$$ So to decrypt, you have to
calculate a discrete logarithm. Given that the size of message space
is small in our experiments, it can be calculated using
``brute-force''. A count value cannot be larger than the duration of
the data collection phase $d$. Even if we control all HSDirs in the
Tor network, the total count cannot be larger than $6\cdot d$. For
further optimization and more efficient calculation of discrete
logarithm, other general approaches such as baby-step
giant-step~\cite{Shanks}, Pollard's kangaroo
algorithm~\cite{Pollard:2000:KMD:2815936.2815952} or Pollard's rho
algorithm~\cite{pollard:1978} can be employed.

Note that secret key $a$ is not known to anybody. Yet, to enable joint
decryption, we make use of the property that each party $i$ knows
share $a_i$, and $a=\prod_i{}a_i$. So, during decryption of ciphertext
$(C_1,C_2)$, party $i$ computes intermediate value $v_i=C_2/C_1^{a_i}$
and sends the result to party $i+1$. Party $i+1$ takes $v_i$ and
computes $v_{i+1}=v_{i}/C_1^{a_{i+1}}$ and so on. Eventually, the last
party publishes the recovered plaintext.

\BfPara{Data Collection.} Each DC stores a simple database of (key,
value) pairs.  When receiving the descriptor of a hidden service, the
DC verifies whether the hash of the .onion address already exist in
the database. If it exists, the DC increments its counter value by
calculating $E_{A}(r, 1) = (g^r, A^r \cdot h)$ and homomorphically
adding $E_{A}(r, 1)$ to the current encrypted count in its
database. If the hash of the .onion address does not exist in the
database, the DC creates a new entry and stores $E_{A}(r, 1)$ for this
hash.  We use an epoch of 24 hours to consider a new hidden
service. After a hidden service is observed by a DC $c$ times, the
value for the counter will consequently be $(g^r, A^r \cdot h^{c})$.

\BfPara{Aggregation.} At the end of an epoch, each DC publishes a
commitment of their encrypted counts to the PBB. After all DCs have
committed, they open their commitment by publishing their encrypted
counts to the PBB, i.e., for each onion address they publish the
encrypted count $ (H(onion), (g^{r}, A^{r} \cdot h^{c} ))$. After
opening commitments, the TKSs verify counts against commitments and if
successful continue operation.

At the end of the data collection phase, one TKS, e.g., TKS$_1$ sums
up all encrypted counts by componentwise multiplication of ciphertexts
belonging to the same $H(onion)$. TKS$_1$ publishes result $R$ which
is a list of ciphertexts $C=E_A(r,c))$. All other TKS re-do this
computation and check TKS$_1$'s $R$. If it matches their computation,
they continue.

%There, each TKS privately aggregates the encrypted counts of their DCs
%by , thus adding the
%underlying plaintexts.

\BfPara{Shuffling.}  The last step is that all TKS engage in a
privacy-preserving shuffle of $R$. Therewith, they will break the
connection between $H(onion)$ and the encrypted count corresponding to
$H(onion)$.

Specifically, TKS$_1$ starts by computing and publishing a random
permutation $R'$ of $R$ where all ciphertexts in $R$ are
re-encrypted. TKS$_2$ computes another random permutation $R''$ of
$R'$ and publishes $R''$, and so on.

To protect against malicious TKSs, we use standard zero-know\-ledge
proofs of knowledge of correctness of shuffles (permutation and
re-encryption). That is, each TKS does not only publish their shuffle,
but publishes their shuffle together with a zero-knowledge proof of
knowledge.  There are several (very efficient) variants of such
zero-knowledge proofs~\cite{Neff,groth,sako}. However, we use a simple
cut-and-choose implementation of a zero-knowledge proof of knowledge
of correctness of a shuffle working as follows.  For a given sequence
$R$ of ciphertexts,
\begin{enumerate}
\item TKS$_i$ computes and
publishes two permutations $R_0$ and $R_1$ with re-encrypted
ciphertexts of $R$.

\item TKS$_i$ also computes the hash $H(R_0, R_1)$, where ``,'' is an unambiguous pairing of inputs.
\item If the first bit of $H(R_0, R_1)$ is $0$, then TKS$_i$ publishes
  details of how to permute $R$ to $R_0$ together with all random
  coins $r$ used during re-encryption.
\item If the first bit of $H(R_0, R_1)$ is $1$, then TKS$_i$ publishes
  permutation from $R$ to $R_1$ together with the random coins.
  
\end{enumerate}
The above four steps are repeated $\ell$ times, resulting in a
zero-know\-ledge proof of knowledge of correctness of a shuffle in the
random oracle model, and with a (statistical) soundness error of
$2^{-\ell}$. A reasonable value of $\ell$ is, e.g., $\ell=40$.

\BfPara{Decryption.} After the last shuffle has been published and its
proof verified, this last shuffle is then
decrypted ciphertext by ciphertext.  As described above, each TKS
partially decrypts each ciphertext. The last TKS publishes the list of
$h^{c}$ values.  Given that each count cannot be larger than the
duration of the study, we use brute-force to calculate the discrete
logarithm, but any of the faster, general algorithms mentioned above
are possible, too.

\subsubsection{Security Analysis and Discussion}
The distributed key generation mechanism guarantees establishing a
public-private key pair against malicious adversaries. Committing and
opening the public part of a key share, or using a zero-knowledge
proof of knowledge as PrivEx, is a standard technique to compute an
unbiased key pair.

For simplicity, we use a $k$-out-of-$n$ approach for our distributed
key generation where $k = n$. While it is easy to implement, it has
several drawbacks. For example, if any of the parties aborts the
protocol, no encrypted data can be decrypted.  As an alternative for
more robustness, one could use $k<n$ approaches. Note, however, that
any multi-party computation protocol requires a honest majority.

To comply with Tor's standard ethics we do not store the raw hidden
services at any point. Each DC calculates the hash of an onion
service, $H(onion)$. This mitigates against leaking the hidden
services address to an adversary without prior knowledge of the onion
address. However, it does not protect against an adversary with prior
knowledge of a hidden service. %One approach would be to use a keyed
%hash at the DCs, where the keys are calculated using any distributed
%maltiparty key generation approach. The TKSs do not need to have the
%knowledge of the key.

Along the same lines as distributed key generation, aggregation also
uses a commit-and-prove approach~\cite{commitprove} to securely
compute sums in the presence of malicious adversaries.

There is, however, still one constraint. A malicious adversary can lie
about their input, i.e., their encrypted counts. That is, a malicious
adversary can publish and prove encryptions of counts that are not
reflecting the really observed counts. In the specific context of our
computations, it is important to point out consequences. For example,
if 2 out of 3 parties are malicious, they could both commit to and
publish only encryptions of 0. As a result, the adversary would learn
a random permutation of the benign party's counts.

In general, it is difficult to protect against adversaries lying about
their input. Yet, in contrast to related work, we discuss several
strategies for mitigation. Besides using differential privacy, we
dedicate Section~\ref{s:protection} to more details.

If we assume adversaries with the only constrain that they do not
cheat about their input, than the above protocol is secure in that an
adversary only learns the histogram (and everything they can derive
from their own view), but nothing else about the other parties' input.

%We protect against a curious adversary, by providing an extension to PrivEx in our protocol. In this variant, we do not have access to the individual counts for a hidden services. Therefore, the results from each TKS cannot be compared to another TKS for an individual hidden service. However, all honest parties would generate the same histogram at the end of the protocol. The shuffling stage takes place after the aggregation phase and before the decryption stage.\\

% Figure~\ref{f:privcount} depicts the diagram of the interaction between DC/TKS/PBB, and the steps described.

% \begin{figure}
% \centering
% \includegraphics[width=0.49\textwidth]{figures/lifespan/TorHS.pdf} %\includegraphics[width=0.99\textwidth]{figures/privcount_diagram.pdf}
% \caption{Diagram of the multi party computation process between TKS and DC.}
% \label{f:privcount}
% \end{figure}

%% file: protection.tex
% \subsection{Differential Privacy}
% {\color{red} \textbf{ @all: Should we provide some discussion on what are the parameters if we wanted to use differential privacy? How? \\
% @Erik: Is deferential privacy in the threat model? What are the advantages or disadvantages compared to shuffling?
% }}

\subsection{Protection Against Dishonest Adversary}\label{s:protection}
In this section, we describe and discuss several first ideas to detect
and mitigate against dishonest parties (DC and TKS) who lie and report
fake counts. Dishonest DCs and TKSs can skew the histograms and final
counts by reporting fake values.

\BfPara{Approach 1.} The rationale behind our first approach is that
parties control each other and verify their output. Due to the random
distribution of descriptor-ids on HSDir relays, there is a non-zero
probability that the same descriptor-id is placed on HSDir relays of
multiple parties. Increasing the number of HSDir relays increases this
probability.

Now let $P_1,\ldots,P_n$ be parties, with at most one being
malicious. Each party controls their own subset of HSDir relays of
size $x$. Parameter $x$ must be chosen carefully in advance such that
it gives certain properties on the precision of the mapping between
counts and hidden service lifetime later.  The protocol goes as
follows. 
\begin{enumerate}

\item Parties $P_1,\ldots,P_n$ enter the data collection phase at the
  same time. During measurement, each party individually and
  independently of other parties counts occurrences of hidden services
  as before.
  
\item At a predetermined deadline, all parties cease measurement. Each
  party $P_i$ computes a histogram $\histo_i$ of counts.

\item Parties engage in an interactive protocol.
  \begin{enumerate}
  \item Each party $P_i$ publishes a commitment  to its histogram $\histo_i$.
    
  \item After all parties have committed, they open commitments and publish their histogram $\histo_i$.
    
  \item For any pair of histograms $\histo_i$ and $\histo_j$, parties
    locally compute histogram \emph{distances}
    $\Dis(\histo_i,\histo_j)$. There are several distances for
    histograms, e.g., Kull\-back-Leib\-ler, $\chi^2$ distance, Earth
    Moving Distance, quadratic form distance, and  combinations
    of that.
    In any case, the output of this step are numbers
    $d_1=\Dis(\histo_1,\histo_2),d_2=\Dis(\histo_1,\histo_3),\ldots,d_{n-1}=\Dis(\histo_{n-1},\histo_{n})$. For
    simplicity, assume $d_1<\ldots{}<d_{n-1}$.
    
  \end{enumerate}
\end{enumerate}
Now, the ``true'' histogram is $\histo_1$.  The rationale is that
histograms of honest, non-malicious parties will be closer than the
histograms of a honest party and a malicious.

Obviously, for this mitigation strategy to work in a straightforward
manner, the number of HSDir relays for each party must be as high as
the number of relays for the set of all parties in the previous
section. However future work could investigate trade-offs between the
choice of $x$ and a compromise between security, cost, and precision
of results.

%ignore

\BfPara{Approach 2.} In this approach, each party creates a set of
``private'' hidden services, i.e., without announcing them to other
parties. By running these hidden services, each party knows exactly
which HSDir relays should have hosted their own descriptors-ids at any
time.  To verify if another party is honest, each party can ask for
the decryption of the count of their hidden service after proving that
they own the hidden service's corresponding private key.

More formally, each party $P_i$ owns $k$ RSA (private, public) key
pairs $(d, e)$ and runs $k$ private hidden services $onion_i$.
During initialization, each party commits to their $k$ onion services.
After data collection, each party then open commitments and proves the
ownership of each of their $onion_i$ using the private key.

After the ownership of a hidden service is verified, each $DC_i$
reveals their count for $onion_i$. If $onion_i$ was online all the
time, everybody knows the exact count of any hidden service for each
DC using the DC's fingerprint.  If count reported from a DC diverges
from its expectation, the DC has cheated. In some sense, this
technique is like a spot checking technique for onion services.

This approach requires that parties actually run their onion services
and not falsely blame other parties. To prove that their services were
really online, one party could reveal an onion service to a subset of
the other parties which would then regularly check availability of the
onion service and serve as witnesses later on.

\BfPara{Approach 3.} Considering that an adversary can detect the
honeypots, another approach is to verify the counts of a set of all
hidden services. This is a variant of the second approach. We sample a
random set of all online hidden services, and query each entity for
their corresponding counts. In this method, the adversary is not able
to report fake values on the known hidden services that do not belong
to the verifiers. Each verifier checks the lifespan of any hidden
service that is uploaded to his HSDir, by performing a non-intrusive
probing. Such probing is carried out by establishing a connection to
the hidden service, without making an HTTP request. Imagine that in a dynamic HSDir circle formation, the probability that an honest HSDir reports a count outside the threshold is $p_1$. Therefore, after $i$ iterations of sampling hidden services lifespan, each TKS can establish a level of confidence on the
correctness of the counts. In this model, each TKS entity can set
their own threshold $\delta$ and only consider the counts that meet
their security requirements.

%% file: implementation.tex
\section{Implementation and Deployment}\label{s:impl}
% \fixme{text from safety board}
To ensure that we are not compromising the security and privacy of Tor, we took the following procedures. We setup a website detailing our project, methodology, fingerprints of our relays, and the identities of the research team. We limit the bandwidth of our relays to 250KB/s, to prevent acquiring the \emph{Guard} flag. Additionally, we disable the egress traffic to the Internet from our relays, to avoid acquiring the \emph{Exit} flag. The \emph{Exit} nodes are specially important component of the Tor network, because they have access to the traffic (if not encrypted). They are also a target of abuse complaints, since their IP is listed as the source of connection to any Internet domain. Furthermore, if the same entity is chosen as both the \emph{Guard} node and \emph{Exit} node, he can perform network correlation attack and de-anonymize Tor users. Therefore, we listed all of our relays as one \emph{Effective Family}, to prohibit two of our relays to be chosen in any one circuit. This ensures that our relays only serve as \emph{Middle} relays and HSDirs inside the Tor network.

% \fixme{end text from safety board}

In the following section, we describe our implementation and deployment of the privacy preserving data collection and analytics. We took the procedures necessary to ensure the security and privacy of the data collected.

\begin{table}
\centering
\scalebox{1}{
\begin{tabular}{ c c c c  }
 \hline
  \textbf{Country} & \textbf{City} & \textbf{Cloud Provider} & \textbf{Count} \\ \hline \\
        USA     &         City 1~\footnote{This city and private cloud are anonymized to satisfy the ACSAC double-blind submission requirement. This information will be revealed in the final version of the paper.}      &         Private      &      20    \\ \hline
        USA   &         Chicago   &       Vultr  &      2   \\ \hline
        USA   &         Dallas    &       Vultr  &      2   \\ \hline
        USA   &         Atlanta   &       Vultr  &      2   \\ \hline
        USA   &         Miami   &       Vultr  &      2   \\ \hline
        Germany &       Frankfurt   &       Vultr  &      6   \\ \hline
        UK    &         London    &       Vultr  &      6   \\ \hline
        France  &       Paris   &     Online SAS   &      32    \\ \hline
        Netherlands &     Amsterdam &     Online SAS   &      8   \\ \hline
\end{tabular}
}
\caption{The distribution of our relays per country, city, and cloud provider. We chose diverse location for the relays, both for security reasons and also to help the Tor network by contributing relays and bandwidth.} 
\label{t:relays_cloud_dist}
\end{table}

We deployed 80 relays that were managed by three different teams, two from the USA and another one from Europe. We divided our relays between the teams at different geolocations for security and privacy reasons. 40 relays (20 VM) were managed by Entity 1, 20 relays (10 VM) managed by Entity 2, and 20 relays (10 VM) by Entity 3. We hosted two relays per VM instance. We used a combination of private and public cloud to host our relays. 20 relays (10 VM) were hosted on private cloud, 20 relays (10 VM) on Vultr cloud provider, and 40 relays (20 VM) on Online SAS. We chose these two cloud providers because of low cost, and that both are in the top 5 cloud providers used by the Tor relays. These providers have data centers in multiple geolocations, allowing us to distribute our relays, both for security reasons and also to help the Tor network. The 20 private cloud relays, were hosted in City 1, USA; 32 Online SAS relays in Paris, France; 8 Online SAS relays in Amsterdam, Netherlands, 6 Vultr relays in Frankfurt, Germany; 6 Vultr relays in London, UK; 2 Vulr relays in Miami, USA; 2 Vultr relays in Atlanta, USA; 2 Vultr relays in Chicago, USA; and 2 Vultr relays in Dallas, USA. Table~\ref{t:relays_cloud_dist} summarizes the distribution of our relays per country, city, and cloud provider. Please note that the count is the number of the relays, not the number of VM instances. There are two relays per VM instance, since only two relays can run on a single IP address, per Tor policies.

We made modifications to the Tor software to log the hash of the hidden services and the time they were uploaded to the HSDirs into an in-memory database. For privacy and security reasons, we used the Redis in-memory database, with the persistence storage disabled. As a result, as soon as the machine is restarted, all unencrypted data that resides in the memory will be deleted. We implemented the privacy preserving counting and aggregation in Python. At the end of each 24 hour epoch, the Python program reads the data from the Redis database, writes the encrypted data to disk, and flushes the databases. At the end of each cycle, we publish the data to the PBB. We used a private git server as our PBB, to collect data from all DCs. When the data collection phase is over, all TKS aggregate the counts, calculate their contributions and publish it to the PBB as well.

Note that the encrypted data on the DCs, cannot be decrypted until all parties participate. The reason behind using a $k$ out of $n$ (where $k = n$) encryption technique is to ensure the security of the collected data. Even if parties are coerced in disclosing their private keys, as long as one party deletes his private key, the encrypted data cannot be decrypted. Moreover, both our DCs and the TKSs reside in different geolocations with different jurisdictions, which makes the coercion of the private keys less practical.

\begin{figure}
\includegraphics[width=0.45\textwidth]{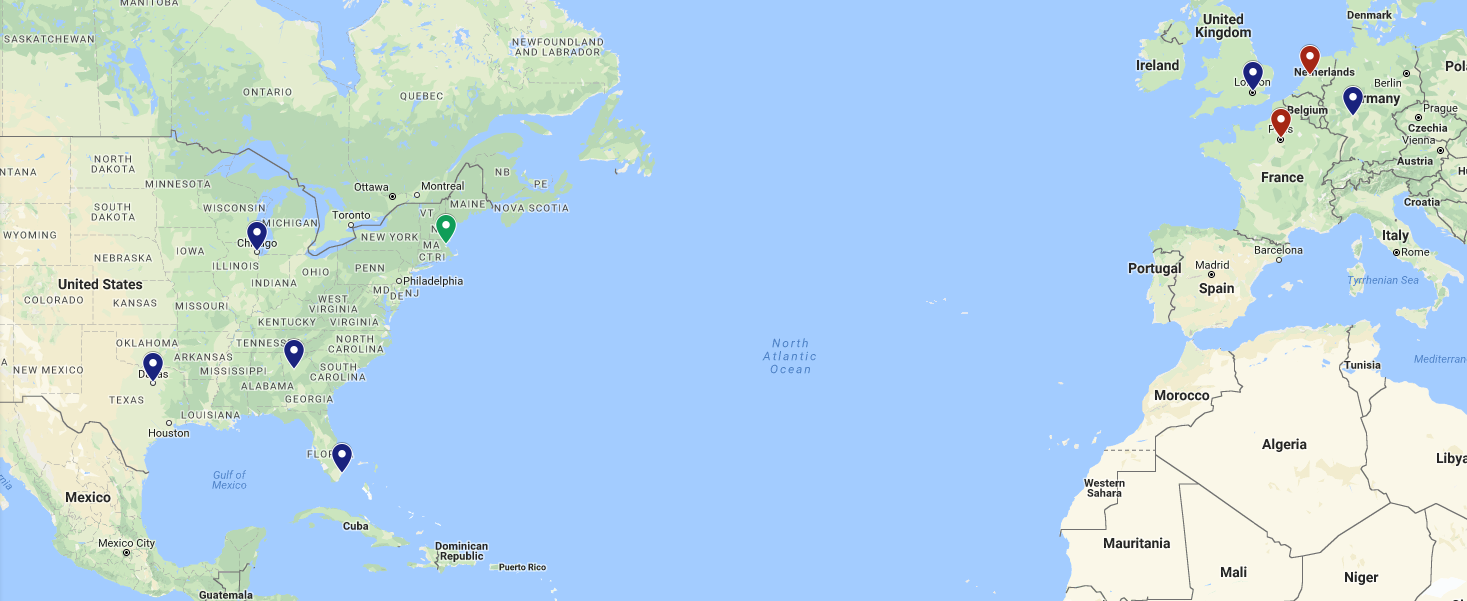}
\caption{The geolocation map of our relays.}
\label{f:relay_map}
\end{figure}

Figure~\ref{f:relay_map} depicts the geolocation distribution of our HSDir relays. As evident, we distributed our servers to different geolocations for security and privacy reasons. Furthermore, we helped the Tor network by contributing relays and bandwidth at different data servers.

%% file: results.tex
\section{Results}\label{s:longresults}

In this section, we discuss the results and findings of our study. As
mentioned previously, we deployed 80 HSDirs over a range of public and
private cloud providers. Our experiments lasted for 180 days, from
September 15, 2017 to March 15, 2018. During this period, we collected
the lifespan counts for 317090 hidden services from our relays.

\begin{figure}

\hspace{1cm}\includegraphics[width=0.45\textwidth]{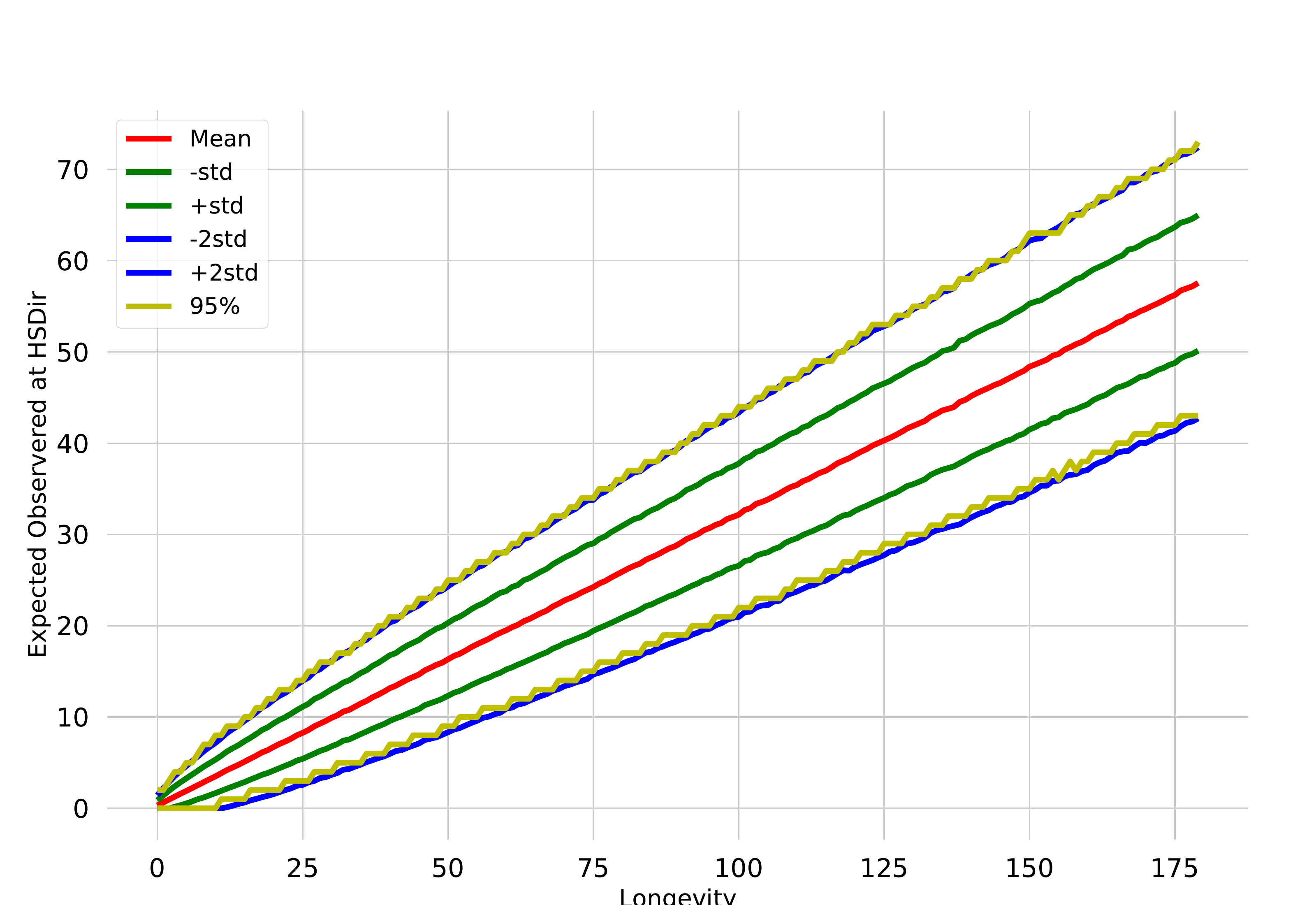}
\caption{Expected counts at HSDirs as a function of lifespan. As we
  can see, 95\% of the expected values are within two standard
  deviation of the mean expected value.}
\label{f:expected_dist}
\end{figure}

We use simulations to find the expected observation counts as a
function of the lifespan of a hidden service for the duration of the
our study. Figure~\ref{f:expected_dist} shows the expected values for
lifespan of hidden services. We calculate the mean, one standard
deviation, two standard deviations and 95\% of the values from the
mean value. The distribution of expected counts follows the bell curve
distribution. As we can see, 95\% of the expected values are within
two standard deviations of the mean expected value.

\begin{figure}
\hspace{0.5cm}\includegraphics[width=0.45\textwidth]{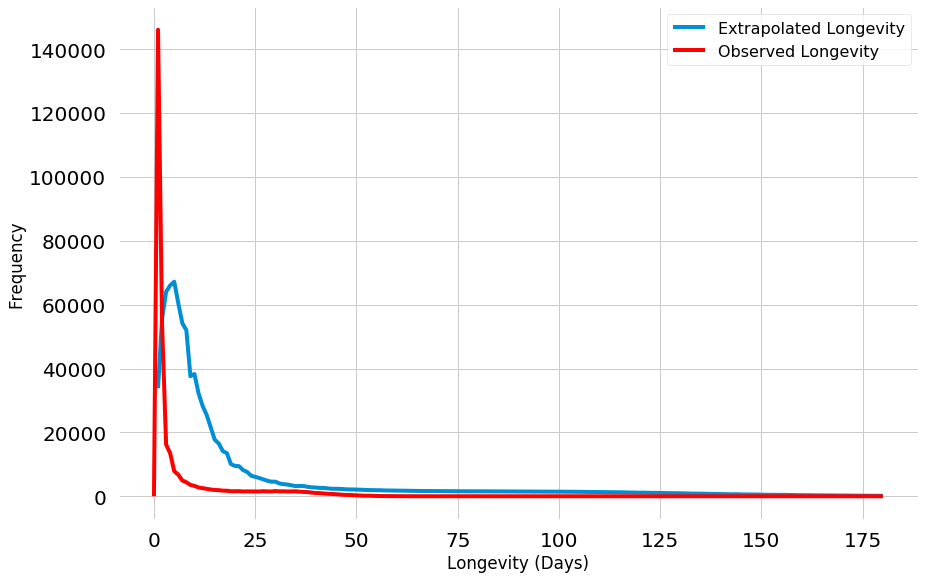}
\caption{Distribution of the estimated lifespan and raw lifespan counts from the HSDirs.}
\label{f:extrapolate_raw_plot}
\end{figure}

Figure~\ref{f:extrapolate_raw_plot} plots the distribution of the
extrapolated lifespan and the raw lifespan counts from the HSDirs. As
we can see around 50\% of the collected hidden services have count of
1. However, in the extrapolated lifespan graph, the majority of the
hidden services have an estimated lifespan of 2 days. This is because
if a hidden service has a count of 1, it can be, with a high
probability, a hidden service with lifespan of 1, 2, 3, or
higher. However, if a hidden service is counted 2 times, it can only
have a minimum lifespan of 2 days. This imbalance between the
estimated lifespan and the observed counts for values less than 10,
explains the different distribution shape between the raw lifespan
counts and the estimated lifespan of hidden services.

\begin{figure}
\hspace{0.5cm}\includegraphics[width=0.45\textwidth]{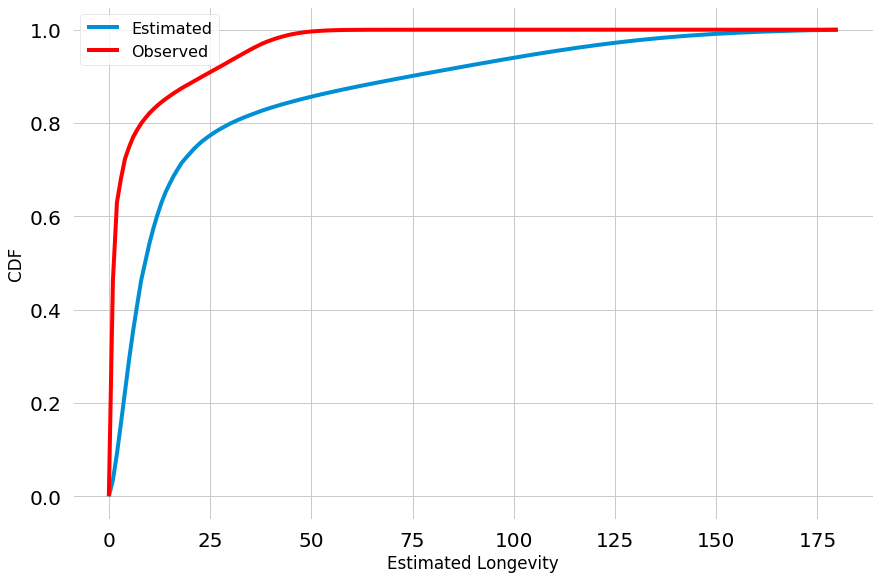}
\caption{CDF of distribution of the estimated lifespan and raw lifespan counts from the HSDirs.}
\label{f:cdf_extrapolate_raw}
\end{figure}

Figure~\ref{f:cdf_extrapolate_raw} depicts the CDF of the observed
counts and the estimated lifespan. More than 60\% of hidden services
are observed 2 times or less on the HSDirs. At the same time, more
than 50\% of the hidden services, have an estimated lifespan of less
than 10 days, where 40\% of them are online for only less than a
week. Our results show that a large majority of hidden services, have
short lifespan, less than a few weeks. Our findings confirms the
results of previous work~\cite{owen2015tor}, regarding the number of
hidden services with short life span. However, the previous work did
not consider any safeguards, and their results were based on the
probing of the collected onion address from their HSDirs, which is
against Tor user agreement. In contrast, our work does not store any
information about the onion services. Furthermore, we do not perform
any probing of the hidden services. Our results rely on the
extrapolation of lifespan based on the privacy-preserving aggregated
counts from a small number of hidden service (2\% of the hidden
services).

Without a holistic view of all the hidden services (i.e., controlling
all HSDirs), it is difficult and non-trivial to draw conclusions on
the churn of the hidden services. However, by using the hidden
services statistics that the Tor project collects from all relays and
the lifespan of hidden services, it can be inferred that the hidden
services have a high churn. A possible explanation for such
short-lived hidden services with high churn can be the dynamics and
behavior of the applications that use hidden services. For example,
OnionShare, a secure and private file sharing that uses Tor hidden
services, creates a new random hidden service address for each file
transfer. As soon as the file is downloaded by the recipients, the
hidden service is discarded.

%% file: related.tex
\section{Related Work}\label{s:longrelated}
% \fixme{Sherr paper}: counts the number of unique items privately. In contrast,
% we compute multiple different counts privately, specifically how often
% each unique item has been counted.

% \fixme{text from intro}
Previous work looking at hidden
services~\cite{owen2015tor}, did not consider any safeguards, and
their results were based on the probing of the collected onion address
from their HSDirs, which is clearly against Tor user agreement. In
contrast, our work does not store any information about the onion
services. Furthermore, we do not perform any probing of the hidden
services.
% \fixme{end text from intro}

In this section, we review the related work on study of Tor hidden services, and advancements in the privacy preserving data collection and analytics.

One of the first studies on the Tor network by McCoy et
al.~\cite{mccoy2008shining}, collected data from the network by
participating and contributing relays to the Tor network. The authors
were studying three main questions: ``How is Tor being used?'', ``How
is Tor being mis-used?'', and ``Who is using Tor?''. Although such
research questions and their answers are important, unfortunately, the
authors did not use the required safeguards to protect the privacy of
users, and mitigate against its compromise. Another example of studies
looking at Tor and hidden services that does not implement security
and privacy safeguards is~\cite{owen2015tor}, where authors collect
information about hidden services by setting up HSDirs. Additionally,
the authors actively probe the collected hidden services for
content. Such work, have been widely criticized by other
researchers~\cite{soghoian2011enforced} and the Tor Project. As a
result tools such as ExperimenTor~\cite{bauer2011experimentor} and
Shadow~\cite{jansen2011shadow} were introduced, for researchers to
experiment with simulated Tor network. As a measure to prevent future
unethical work, the Tor Research Safety Board~\cite{torboard} was
established to review the research projects studying the Tor network,
to ensure such works do not harm users' privacy. In the next section
we will further discuss the Tor Research Safety Board.

One of the early works on privacy preserving data collection and study
of the Tor network, specially hidden services, is a work by the Tor
team~\cite{goulet2015hidden} to collect and count the number of unique
hidden services. The authors use differential privacy, controlled
noise addition to the statistics, and limiting accuracy to a certain
granularity via binning. However, this work does not study the
lifespan, or the nature of hidden services and how they are used. This
is mainly because reporting such statistics, might harm the privacy of
individual Tor users.

The closest study to our work is PrivEx~\cite{Elahi14}, which our
privacy preserving counting protocol is based upon. In this work, the
authors introduce two variants for secure counting. One with a shared
secret key and another one that relies on multi-party computation. Our
work uses the second variant. The authors implement and deploy their
protocol to collect and study egress traffic from Tor, and case study
the popularity of certain Internet website in different
locations. Such work can be used to investigate the censorship of
content in different locations, in a privacy preserving
manner. PrivCount~\cite{Jansen:2016:SMT:2976749.2978310} is another
work that relies on PrivEx, for measuring the Tor network designed
with user privacy as a primary goal. It aggregates the collected data
from different Tor relays to produce differentially private
outputs. Another work that uses shuffling of the data to hide
relations between the elements and counts is
PSC~\cite{fenske2017distributed}, which is a work on private set union
cardinality~\cite{7961247}.

Another work~\cite{jansen2017inside} uses Tor circuit and website
fingerprinting techniques from the middle relays to detect the hidden
services popularity and usage. Furthermore, the authors use their
fingerprinting technique to study the popularity of a social
networking hidden service, by setting up middle relays in the Tor
network. The authors demonstrate the possibility of such passive study
in a privacy preserving manner from the middle relays. In another
work~\cite{histore}, the authors introduce Histor, which relies on
differential privacy to collect and analyze information about Tor
statistics.

%Machine learning
Privacy preserving techniques are also being used in machine learning
and deep learning algorithms for secure aggregation of
high-dimensional data. For example,
~\cite{Bonawitz:2017:PSA:3133956.3133982} allows the computation of
sum of large values at the honest-but-curious and active adversary
server, using the data from different clients. Dolev et
al.~\cite{dolev2016private} introduce privacy preserving algorithms
for computation and performing search, fetch, and range queries with
the map-reduce framework. Related to our work is also
Prio~\cite{Corrigan-Gibbs:2017:PPR:3154630.3154652}, a
privacy-preserving system for the collection of aggregate statistics.

%% file: conclusion.tex
\section{Conclusion}\label{s:conclusion}

Tor and hidden services attracted a large user base in the past few
years. Investigating the hidden services dynamics and pro-actively
protecting them against abuse and subversion is crucial in their
success. Given the strict privacy requirement and security assumption
of Tor, it is important to consider the implication of data collection
and analysis in such infrastructure. In this paper, we demonstrate
that it is possible to carry-out a privacy-preserving study of hidden
services longevity, satisfying the Tor safety board requirements, yet
obtaining fairly accurate estimations.